\documentclass[12pt]{article}
\usepackage{graphicx}
\usepackage{amssymb}
\usepackage{amsmath}

\newcommand{\gsim}{ \mathop{}_{\textstyle \sim}^{\textstyle >} }
\newcommand{\lsim}{ \mathop{}_{\textstyle \sim}^{\textstyle <} }
\newcommand{\bear}{\begin{array}}  
\newcommand {\eear}{\end{array}}
\newcommand{\bea}{\begin{eqnarray}}   
\newcommand{\eea}{\end{eqnarray}}
\newcommand{\beq}{\begin{eqnarray}}   
\newcommand{\eeq}{\end{eqnarray}}
\newcommand{\bef}{\begin{figure}}  \newcommand 
{\eef}{\end{figure}}
\newcommand{\bec}{\begin{center}}  \newcommand 
{\eec}{\end{center}}
  
\newcommand{\la}{\left\langle}  
\newcommand{\ra}{\right\rangle}

\newcommand{\meg}{\mu \rightarrow e \gamma}
\newcommand{\MSUSY}{M_{\text{SUSY}}}

\newcommand{\tb}{\tan{\beta}}
\newcommand{\sa}{\sin{\alpha}}
\newcommand{\ca}{\cos{\alpha}}
\renewcommand{\sb}{\sin{\beta}}
\newcommand{\cb}{\cos{\beta}}
\newcommand{\sab}{\sin{(\alpha - \beta)}}
\newcommand{\cab}{\cos{(\alpha - \beta)}}

\newcommand{\mNeN}{\mu N \rightarrow e N }
\newcommand{\meee}{\mu \rightarrow 3e}
\newcommand{\BR}{\text{BR}}

\begin{document}

\begin{titlepage}

\begin{flushright}
IPMU~10-0086\\
ICRR-Report-569-2010-2\\
MISC-2010-03
\end{flushright}

\vskip 1.35cm

\begin{center}

{\large Reevaluation of Higgs-Mediated $\mu$-$e$ Transition in the MSSM }

\vskip 1.2cm

Junji Hisano$^{a,b,c}$, 
Shohei Sugiyama$^{a,c}$,
Masato Yamanaka$^{a,d}$,\\
and Masaki Jung Soo Yang$^{a,c}$,

\vskip 0.4cm

{\it $^a$Institute for Cosmic Ray Research,
University of Tokyo, Kashiwa 277-8582, Japan}\\
{\it $^b$Institute for the Physics and Mathematics of the Universe,
University of Tokyo, Kashiwa 277-8568, Japan}\\
{\it $^c$Department of Physics, Nagoya University, Nagoya 464-8602, Japan}\\
{\it $^d$Maskawa Institute for Science and Culture, Kyoto Sangyo University,
Motoyama, Kamigamo, Kita-Ku,Kyoto, 603-8555, Japan}

\date{\today}

\begin{abstract} 
  In this letter, we reevaluated the Higgs-mediated contribution to
  $\mu \to e \gamma$, $\mu \to 3 e$, and $\mu$-$e$ conversion in
  nuclei in the MSSM, assuming left-handed sleptons have flavor-mixing
  mass terms. Contrary to previous works, it is found that Barr--Zee
  diagrams including top quark give dominant contribution to $\mu \to
  e \gamma$, and those including bottom quark and tau lepton are also 
  non-negligible only when $\tan\beta$ is large.  As a result, the
  Higgs-mediated contribution dominates over the gaugino-mediated
  contribution at one-loop level in $\mu\rightarrow e \gamma$ when
  $\MSUSY/m_{A^0}\gsim 50 $, irrespectively of $\tan\beta$ as far as
  $\tan\beta$ is not large.  Here, $\MSUSY$ and $m_{A^0}$ are a
  typical mass scale of the SUSY particles and the CP-odd Higgs boson
  mass, respectively. Ratio of branching ratios for $\mu\rightarrow e
  \gamma$ and $\mu$-$e$ conversion in nuclei is also evaluated by
  including both the gaugino- and Higgs-mediated contributions to the
  processes. It is found that the ratio is sensitive to $\tan\beta$
  and $\MSUSY/m_{A^0}$ when $\MSUSY/m_{A^0}\sim (10-50)$ and
  $\tan\beta\gsim 10$.

\end{abstract} 

\end{center}
\end{titlepage}

\section{Introduction}

Charged lepton-flavor violating (cLFV) processes, such as
$\mu\rightarrow e \gamma$, are sensitive to physics beyond the
standard model (SM) \cite{Raidal:2008jk}. While the lepton-flavor
conservation is not exact in nature due to finite neutrino masses, 
cLFV processes are quite suppressed in the standard model. Thus,
searches for cLFV processes are a unique window to physics beyond the
SM, especially, at TeV scale.

Now the MEG experiment is searching for $\mu\rightarrow e \gamma$
\cite{meg}, it would reach to $\sim 10^{-13}$ for the branching ratio
on the first stage, which is improvement of two orders of magnitude
compared with the current bound. The COMET and Mu2e experiments
\cite{mu2e,comet}, which are searches for $\mu$-$e$ conversion in
nuclei, are being planed in J-PARC and Fermilab, respectively. It is
argued that they would reach to $\sim 10^{-16}$ for branching ratio of
$\mu$-$e$ conversion with target $\rm Al$. Here, branching ratio of
$\mu$-$e$ conversion is ratio of $\mu$-$e$ conversion rate over muon
capture rate. Searches for $\mu\rightarrow e \gamma$ and $\mu$-$e$
conversion in nuclei are complementary to each other in studies of
physics beyond the SM since those processes may be induced by
different types of processes.

The minimal supersymmetric (SUSY) standard model (MSSM) is a leading
candidate for physics beyond the SM, and cLFV processes are
extensively studied in the model. SUSY-breaking slepton mass terms are
lepton-flavor violating.  It is noticeable that ratios of branching
ratios for cLFV processes would give information of mass spectrum in
the MSSM, since dominant diagrams in cLFV processes depend on
the mass spectrum.

When SUSY particle masses are $\lsim O(1)$~TeV, the muon LFV processes,
such as $\mu$-$e$ transition processes, $\mu\rightarrow e \gamma$,
$\mu\rightarrow 3 e$ and $\mu$-$e$ conversion in nuclei, are
generated by the gaugino-mediated contribution, which is generated by
one-loop diagrams of gauginos and sleptons (and Higgsinos).  Branching
ratios for the cLFV processes due to the gaugino-mediated contribution are
suppressed by $1/\MSUSY^4$, since the effective dipole interaction is
dominant in the cLFV processes. Here, $\MSUSY$ is a typical mass scale of
the SUSY particles. On the other hand, when $\MSUSY \gsim O(1)$~TeV,
the Higgs-mediated contribution to the processes could be
sizable. The non-holomorphic LFV correction is generated to Yukawa
coupling of the Higgs bosons at one-loop level, and it is not
suppressed by $\MSUSY$ \cite{Babu:2002et}.  Branching ratio of
$\mu$-$e$ conversion in nuclei is more sensitive to the Higgs-mediated
contribution \cite{Kitano:2003wn}. Thus, ratio of branching
ratios for $\mu\rightarrow e \gamma$ and $\mu$-$e$ conversion in
nuclei is a good observable to constrain mass spectrum in the MSSM,
since it is sensitive to whether the gaugino-mediated or
Higgs-mediated contribution is dominant.

In this letter we systematically calculate the Higgs-mediated contributions 
to cLFV reactions in the MSSM, and clarify the dominant process in each 
cLFV reaction.
For this purpose, we first evaluate the Higgs-mediated
contribution to $\mu\rightarrow e\gamma$ in the MSSM. Barr--Zee
diagrams give dominant contribution to $\mu\rightarrow e \gamma$ among
various diagrams though those are of higher order. We systematically
evaluate those diagrams, and find that Barr--Zee diagrams including
top quark give the largest contribution, and the branching ratio for
$\mu\rightarrow e \gamma$ induced by the Barr--Zee diagrams is
approximately proportional to $\tan^2\beta$. The angle $\beta$ is
defined by $\tan \beta = { \la H_{2}^{0} \ra / \la H_{1}^{0} \ra } $.
The Higgs-mediated contribution dominates over the gaugino-mediated
one in $\mu\rightarrow e \gamma$ when $\MSUSY/m_{A^0}\gsim 50 $, which
is almost insensitive to $\tan\beta$. Here, $m_{A^0}$ is the CP-odd Higgs
boson mass in the MSSM.

Using this result, we evaluate ratio of branching ratios for
$\mu\rightarrow e \gamma$ and $\mu$-$e$ conversion in nuclei.  When
the Higgs-mediated contribution is dominant, the ratio of the
branching ratios is scaled by $\tan^4\beta$. It is found that the
ratio is quite sensitive to $\tan\beta$ and $\MSUSY/m_{A^0}$ when
$\MSUSY/m_{A^0}\sim (10-50)$ and $\tan\beta\gsim 10$. We also check
that the ratio of the branching ratios for $\mu\rightarrow e \gamma$
and $\mu\rightarrow 3 e$ is insensitive to them.

In Refs.~\cite{Paradisi:2006jp, Paradisi:2005tk} the Higgs-mediated
contribution to the $\mu$-$e$ transition processes in the MSSM is
discussed. It is argued that when the Higgs-mediated contribution is
dominant, the Barr--Zee diagram including $W$ boson is dominant and
branching ratio of $\mu\rightarrow e \gamma$ is scaled by
$\tan^4\beta$, not $\tan^2\beta$.  This is obviously overestimated. We
clarify what is wrong in their deviation.

We assume that left-handed sleptons have flavor-mixing mass terms
in this letter, simply because this setup is well-motivated from the
SUSY seesaw model \cite{seesaw}. Extension to more general cases will
be given elsewhere.

This letter is organized as follows. In the next section we evaluate the
Higgs-mediated contribution for $\mu\rightarrow e \gamma$. We show
ratio of the Higgs-mediated and gaugino-mediated contributions. In
Sec.~3, we discuss the Higgs-mediated contributions to
$\mu\rightarrow 3 e$ and $\mu$-$e$ conversion in nuclei, and evaluate
ratios among the cLFV processes.  Sec.~4 is devoted to
conclusions and discussion.

\section{Higgs-mediated contribution to $\mu\rightarrow e \gamma$ in
  the MSSM }

In the MSSM, LFV in the Higgs coupling originates from the
non-holomorphic correction to Yukawa interaction of charged leptons
\cite{Babu:2002et}. By including the correction due to one-loop diagrams
of gaugino and sleptons, the effective Yukawa coupling is given as
follows:
\begin{equation}
\begin{split}
   \mathcal{-L}_{\rm{eff}} 
   = \overline{e}_{R i}' y_{e i} H_1^{0} e_{L i}' 
   + \overline{e}_{R i}' y_{e i}
   \left( \epsilon_1^{(i)} \delta_{ij} +   {\epsilon}_2^{(ij)}  \right) 
   H_2^{0 *} e_{Lj}' + {\rm h.c.}, 
\end{split}   
\end{equation} 
where $y_{e i}$ stands for the $i$-th charged-lepton Yukawa coupling
constant at tree level and $e_{Ri}'$ and $e_{Li}'$ represent
right-handed and left-handed leptons, respectively, in a basis where
the tree-level lepton Yukawa matrix is diagonal. The non-holomorphic
interaction ${\epsilon}_2^{(ij)}$ $(i\ne j)$ is generated by
flavor-violating slepton mass terms. As mentioned in introduction, we
assumed that left-handed sleptons have flavor-violating mass terms.
We parametrize ${\epsilon}_2^{(ij)}$ with mass insertions (MIs)
parameters, $\delta^{LL}_{ij} = (\Delta
m^2_{\tilde{l}_L})_{ij}/\tilde{m}^2_{\tilde{l}_L}$, where $ (\Delta
m^2_{\tilde{l}_L})_{ij}$ is off-diagonal element of left-handed
slepton mass matrix and $\tilde{m}_{\tilde{l}_L}$ is an average
left-handed slepton mass.  When the SUSY-breaking mass parameters in
the MSSM are taken to be a common value ($M_{\rm SUSY}$), the
non-holomorphic corrections $\epsilon_1^{(i)}$ and
${\epsilon}_2^{(ij)}$ are reduced to
\begin{eqnarray}
\epsilon_1^{(i)} 
&=&
\frac{g_Y^2}{64\pi^2}
-\frac{3g_2^2}{64\pi^2} \ ,
\nonumber\\
{\epsilon}_2^{(ij)}
&=&
\left(-\frac{g_Y^2}{64\pi^2}
+\frac{g_2^2}{64\pi^2}\right)\delta^{LL}_{ij} \ .
\end{eqnarray}
Note that $\epsilon_1^{(i)}$ and ${\epsilon_2}^{(ij)} $ 
do not vanish even in a limit of large masses of SUSY particles.  This is
quite different from LFV effective dipole operators induced by the
gaugino-slepton loops, whose coefficients are suppressed by masses of
internal SUSY particles.

In a mass-eigenstate basis for both leptons and Higgs bosons,
$\mathcal{L}_{\rm eff}$ for $\mu$-$e$ transition is described
as~\cite{Babu:2002et}
\begin{eqnarray}
-{\cal L}_{\rm eff}^{\mu{\text -}e} =
\frac{m_\mu \Delta_{\mu e}^L}{v \cos^2 \beta}
( \bar{\mu} P_L e ) 
\left[
\cos (\alpha - \beta ) h^0
+ \sin (\alpha - \beta ) H^0
- i A^0
\right] + {\rm h.c.}\ ,
\label{Yukawa2}
\end{eqnarray}
where $h^0$ and $H^0$ are the CP-even Higgs fields ($m_{h^0} <
m_{H^0}$), and $A^0$ is the CP-odd Higgs field. The LFV parameter
$\Delta^L_{\mu e}$ is given by $ \Delta^L_{\mu e} = {\epsilon}_2^{(\mu
  e)} / ( 1 + \epsilon_1^{(\mu)} \tan \beta )^2$.  When we treat
$\epsilon_{1}^{(\mu)}$ and $\epsilon_{2}^{(\mu e)}$ as a perturbation,
we may neglect $\epsilon_{1}^{(\mu)}$ of the denominator at the first
order. In this letter, we set $\Delta_{\mu e}^L = {\epsilon}_2^{(\mu
  e)}$ .

In the MSSM, LFV interaction of $h^0$ in Eq.~(\ref{Yukawa2}) vanishes
when the masses of $H^0$ and $A^0$ go to infinity, since
$\cos(\alpha-\beta)$ behaves as
\begin{equation}
\cab \sim { -2 m_{Z^0}^2 \over m_{A^0}^2 \tan\beta}.  
\end{equation}
This comes from a fact that SM does not have LFV and the light Higgs
boson $h^0$ becomes SM-like in above limit.  Therefore the
contributions from $H^0$ and $A^0$ should be included in cLFV
processes.

\begin{figure}[h]
\begin{center}
\begin{tabular}{ccc}
   \includegraphics[width=7cm,clip]{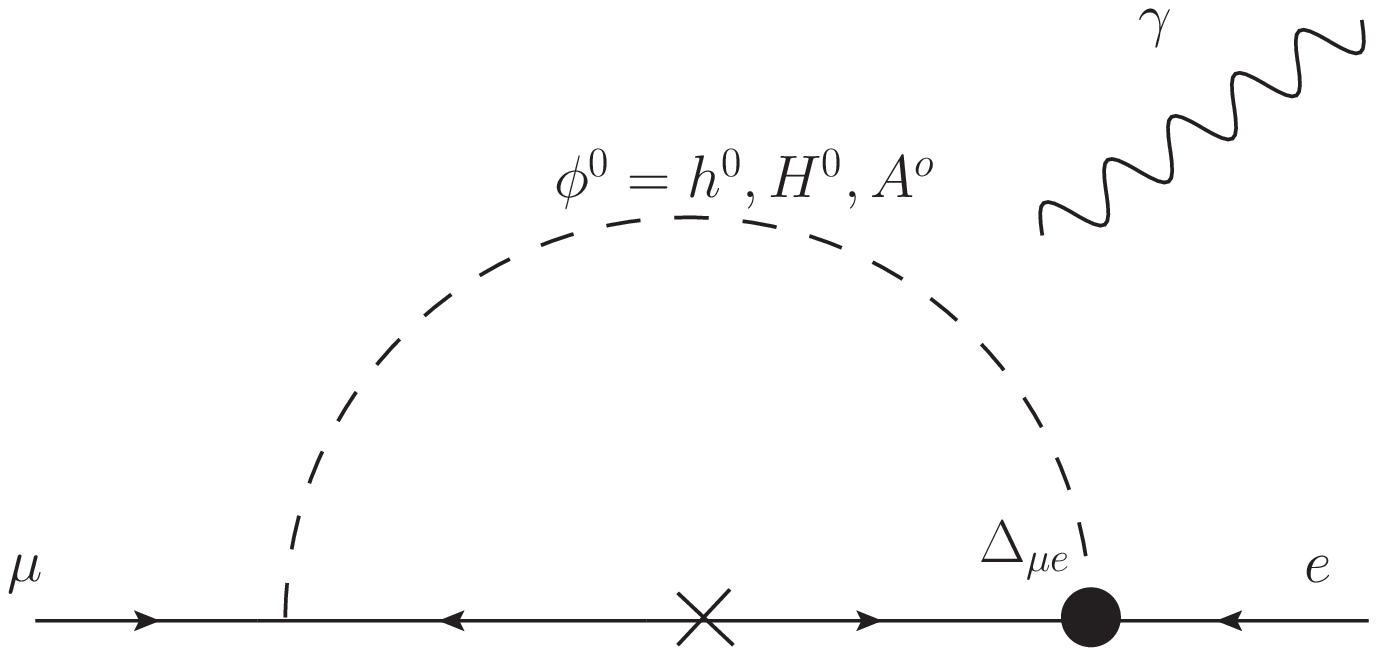} &
   \includegraphics[width=7cm,clip]{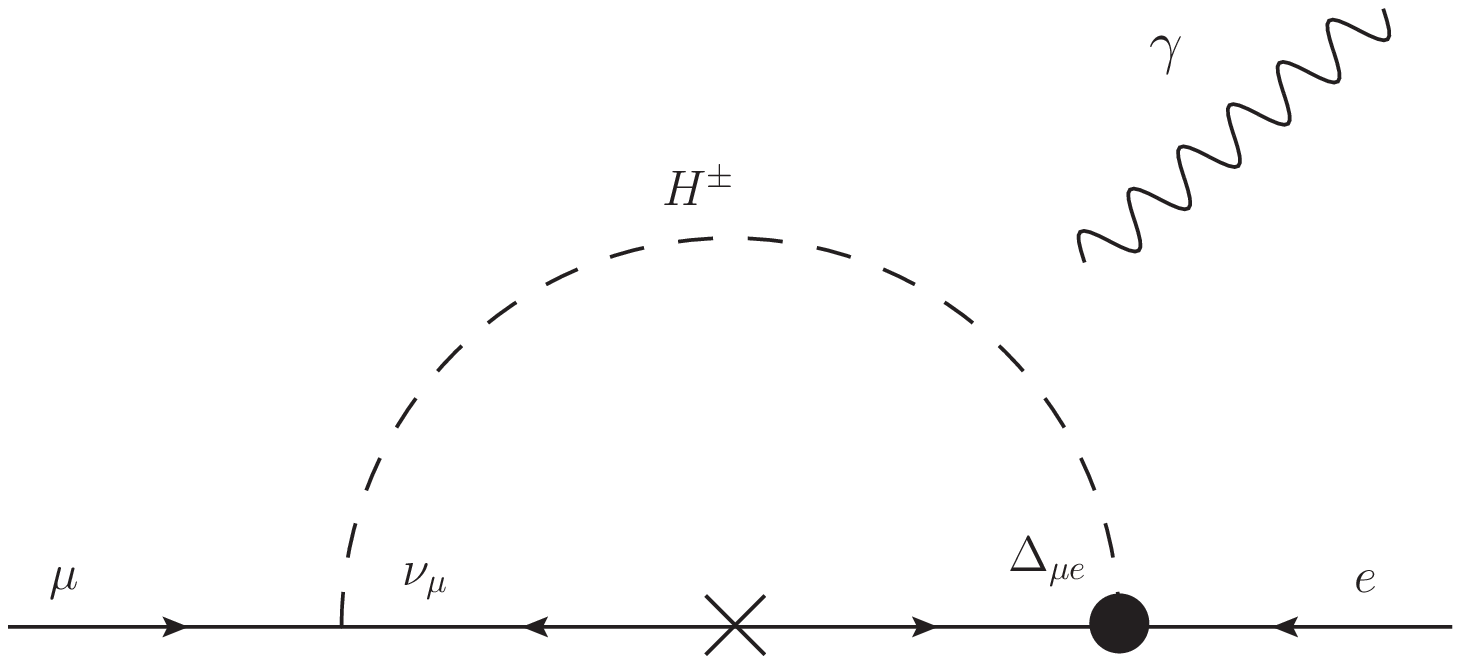} \\
   (a) & (b) 
\end{tabular} 
\caption{$\mu\rightarrow e \gamma$ induced by Higgs boson exchange at one loop level.}
\label{1loop}
\end{center}
\end{figure}

Now we consider the Higgs-mediated contribution to $\mu\rightarrow e
\gamma$ in the MSSM.  Effective amplitude for $\mu\rightarrow e
\gamma$ is parametrized as
\begin{equation}   \label{T}
   T 
   = e \epsilon^{* \mu}(q) ~ \overline{u}_e (p - q) 
   \biggl[ m_{\mu} i \sigma_{\mu \nu} q^{\nu} 
   (A^L P_L + A^R P_R) \biggr] u_\mu (p) \ ,
\end{equation}    
and branching ratio of $\mu\rightarrow e \gamma$ is derived as ${\rm
  BR}(\mu \rightarrow e \gamma) = { (48 \pi^3 \alpha_{\rm em}/ G_F^2)}
(|A^L|^2 + |A^R|^2)$. Here, $\alpha_{\rm em}(\equiv e^2/4 \pi)$ is the
fine structure constant and $G_F$ is the Fermi constant.  While this
amplitude could be induced at one-loop level (Fig.~\ref{1loop}), it is
suppressed by three chiral flips, {\it i.e.}, one chirality flip in
the lepton propagator and two lepton Yukawa couplings. Indeed two-loop
diagrams may be significant contribution. As shown in Fig.~\ref{Barr
  Zee}, two-loop diagrams, called as Barr--Zee diagrams, involve only
one chiral flip (from lepton Yukawa coupling), and hence their
contribution is much larger than that at one-loop level.

\begin{figure}[h]
\begin{center}
\begin{tabular}{ccc}
   \includegraphics[width=7cm,clip]{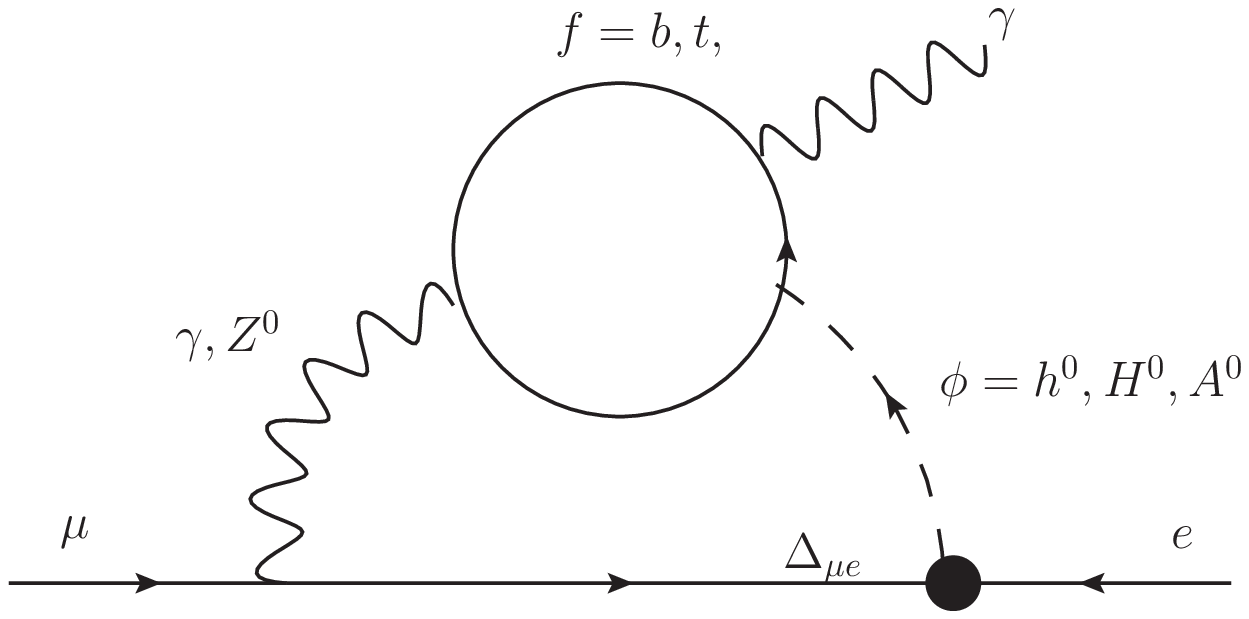} &
   \includegraphics[width=7cm,clip]{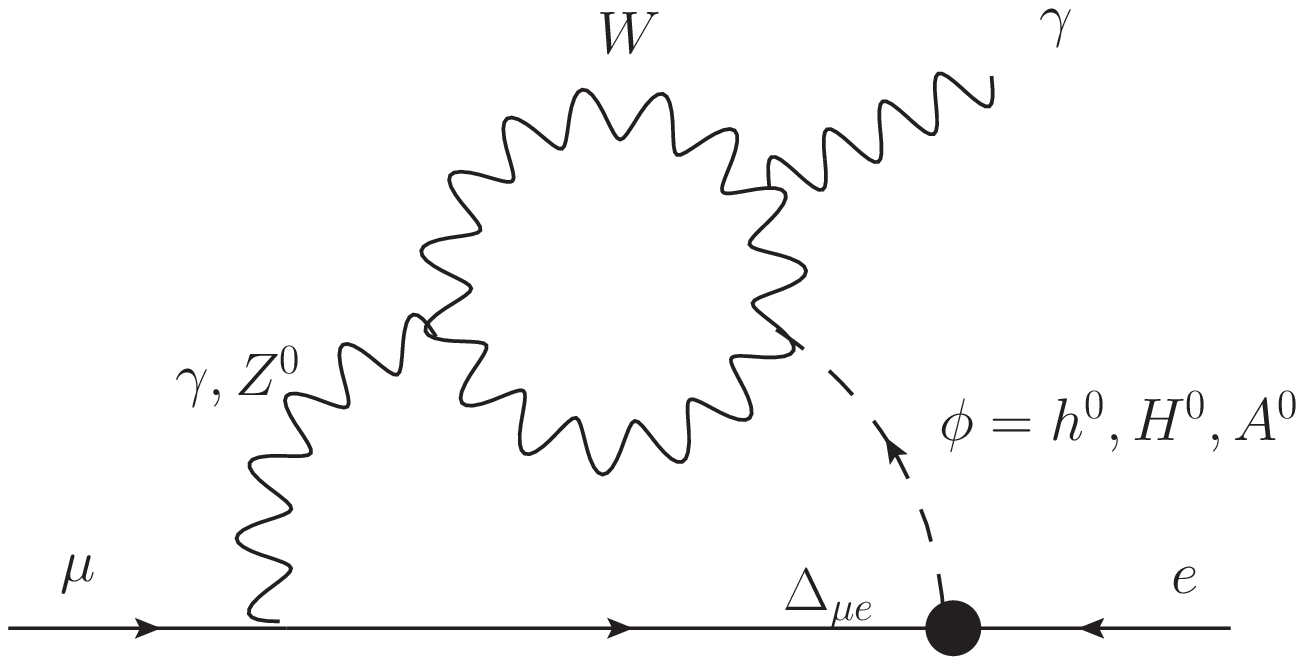} \\
   (a) & (b) 
\end{tabular} 
\caption{Examples of two-loop Barr--Zee diagrams induced by Higgs exchange.}
\label{Barr Zee}
\end{center}
\end{figure}

Following Refs.~\cite{Paradisi:2006jp, Paradisi:2005tk}, we consider
Barr--Zee diagrams which involve effective
$\gamma$-$\gamma$-$\phi^0$ vertices ($\phi^0=h^0$, $H^0$, and $A^0$). The
effective vertices are induced by heavy fermion or weak gauge/Higgs boson
loops.  Barr--Zee diagrams involving bottom- and top-quark loops
(Fig.~\ref{Barr Zee}~(a)) give contributions to the coefficient $A^R$
in Eq.~(\ref{T})  as
\begin{equation}
\begin{split}
   A^R_{{\rm BZ}(b)}  
   &=  {2 \sqrt{2}G_{F}  \alpha_{\rm em} N_c Q_{b}^{2} \over 16\pi^{3} } 
   \Delta^{L}_{\mu e}  \\
   &\times \left[ - {\cab \sa \over \cos^{3}\beta }  
   f(z_{h^0}^{b}) + {\sab \ca \over \cos^{3}\beta }  f(z_{H^0}^{b}) 
   + {\sb \over \cos^{3} \beta}  g(z_{A^0}^{b}) \right] , \\
   A^R_{{\rm BZ}(t)}  
   &=  {2 \sqrt{2}G_{F} \alpha_{\rm em} N_c Q_{t}^{2} \over 16\pi^{3} } 
   \Delta^{L}_{\mu e}   \\
   &\times \left[ {\cab \ca \over \cos^{2}\beta \sb}  
   f(z_{h^0}^{t}) +  {\sab \sa \over \cos^{2}\beta \sb} f(z_{H^0}^{t}) 
   + {1 \over \sb \cb} g(z_{A^0}^{t}) \right] . \\
\end{split}   \label{Abt}
\end{equation}
Here, $N_c$ is color factor, and $Q_{b(t)}$ represents electric
charge for bottom (top) quark. $z_{\phi^0}^{q} =
m_{q}^{2}/m_{\phi^0}^{2}$ for $\phi^0 = h^{0}$, $H^{0}$, $A^{0}$ and
$q = b$, $t$.  Similarly, we calculate the coefficient for tau-lepton
loop by substituting $N_{c} =1$, and replacing $Q_{b}, m_{b}$ to
$Q_{\tau}, m_{\tau}$.  The functions $f(z)$ and $g(z)$ are called
Barr--Zee integrals, whose explicit forms and asymptotic behaviors are
given in Appendix.  For $m_{A^0}\gg m_{Z^0}$ and $\tan \beta \gg 1$,
the Barr--Zee diagram contribution is approximated as
\begin{equation}
\begin{split}
   A_{{\rm BZ}(t,b,\tau)}^R 
   & \simeq {\sqrt{2} G_{F} N_{c} \alpha_{\rm em} \over 8 \pi^{3}} 
   \Delta^{L}_{\mu e} \\
   &~~\times 
   \left[ 
   { Q_t^2 m_t^2 \over m_{A^0}^2 } 
   \tb \left(\log{m_t^2 \over m_{A^0}^2 } \right)^2
   - { Q_b^2 m_b^2 \over m_{A^0}^2 } \tan^3 \beta 
   \left( \log{m_b^2 \over m_{A^0}^2 } + 2\right) 
   - (b \rightarrow \tau) \right] \ . 
\end{split}
\end{equation}
It is found that $\tan\beta$ and/or large logarithmic factors enhance
heavy-Higgs ($H^{0}$, $A^{0}$) contributions, and the light-Higgs
($h^{0}$) contribution is subdominant.

Similarly, the Barr--Zee contributions from loops of $W^-$ boson
(Fig.~\ref{Barr Zee}~(b)), Nambu--Goldstone (NG) boson $G^-$, and charged
Higgs boson $H^-$ are calculated.  Each contribution to $A^R$ is
derived as follows,
\begin{equation}
\begin{split}
   A^{R}_{{\rm BZ}(W^-)} 
   &= { \alpha_{\rm em} \sqrt{2} G_{F} \over 16 \pi^{3}}~ 
   {\sin(\alpha - \beta) \cos(\alpha - \beta) \over \cos^{2} \beta}~ 
      \left[F(z_{h^0}^{W^-})- F(z_{H^0}^{W^-})\right]~  \Delta_{\mu e}^{L}\ ,
\end{split}
\end{equation}
\begin{equation}
\begin{split}
   A^{R}_{\rm BZ(G^-)} 
   &=  {- \alpha_{\rm em} \sqrt{2} G_{F} \over 16 \pi^{3}}~ 
   {\sin(\alpha - \beta) \cos(\alpha - \beta) \over \cos^{2} \beta}~ 
      \left[F'(z_{h^0}^{W^-})- F'(z_{H^0}^{W^-})\right]~  \Delta_{\mu e}^{L} \ , 
\end{split}
\end{equation}
\begin{equation}
\begin{split}
   A^{R}_{{\rm BZ}(H^-)} 
   &= {\alpha_{\rm em} \sqrt{2} G_{F} \over 16 \pi^{3}}~{1 \over \cos^{2} \beta}~
   \left[ {\cos(\alpha - \beta) \over m_{h^{0}}^{2}} f_{h^0} F'(z^{H^{-}}_{h^0}) 
   + {\sin(\alpha - \beta) \over m_{H^{0}}^{2}} f_{H^0} F'(z^{H^{-}}_{H^0}) \right]~ 
   \Delta_{\mu e}^{L}\ , 
\end{split}
\end{equation}
where $z_{\phi^0}^{\phi^-} = m_{\phi^-}^{2} / m_{\phi^0}^{2}$
($\phi^-=W^-, H^-$ and  $\phi^0=h^0, H^0$),
and $f_{\phi^0}$ $(\phi^0 = h^{0},H^{0})$ comes from
coupling of $H^{+} H^{-} \phi^{0}$,
\begin{equation}
\begin{split}
f_{h^0} &= -2 m_{W^{-}}^{2} \sin(\alpha - \beta) + m_{Z^{0}}^{2} \sin{(\alpha + \beta)} \cos{2\beta} \ , \\
f_{H^0} &= 2 m_{W^{-}}^{2}  \cos(\alpha - \beta) - m_{Z^{0}}^{2}  \cos{(\alpha + \beta)} \cos{2 \beta} \ . \\
\end{split}
\end{equation}
The functions $F(z)$ and $F'(z)$ are $3 f(z) + 5 g(z) + {3/4}
g(z)+{3/4} h(z)$ and $(g(z) -f (z))/(2z)$, respectively. $h(z)$ is
also given in Appendix.  The CP-odd Higgs boson $A^{0}$ does not
appear here if CP is conserved.  The above 
contributions are approximated in a limit of $m_{A^0}\gg m_{Z^0}$ and
$\tan \beta \gg 1$ as
\begin{equation}
\begin{split}
   A^{R}_{{\rm BZ}(W^-)} &\simeq { \alpha_{\rm em} \sqrt{2} G_{F} \over 16 \pi^{3}} 
 {2m_{Z^{0}}^{2} \over m_{A^{0}}^{2} } \tan \beta  \Delta_{\mu e}^{L} \\
   &\times \left[ F(\cos^{2} \theta_{W}) - {35 \over 8} {m_{W^{-}}^{2} 
   \over m_{A^{0}}^{2} }  \left(\log{m_{W^{-}}^2 \over m_{A^0}^2 } \right)^2 \right] \ ,
\end{split} \label{bosonicapro}
\end{equation}
\begin{equation}
\begin{split}
   A^{R}_{{\rm BZ}(G^-)} &\simeq { - \alpha_{\rm em} \sqrt{2} G_{F} \over 16 \pi^{3}} 
    {2m_{Z^{0}}^{2} \over m_{A^{0}}^{2} } \tan \beta  \Delta_{\mu e}^{L} \\
   &\times \left[ F'(\cos^{2} \theta_{W}) +{1\over2} \left( \log{m_{W^{-}}^2 \over m_{A^0}^2 } + 2 \right) \right] \ ,
\end{split} \label{bosonicapro2}
\end{equation}
\begin{equation}
\begin{split}
   A^{R}_{{\rm BZ}(H^-)} 
   &\simeq { \alpha_{\rm em} \sqrt{2} G_{F}  \over 16 \pi^{3}} {m_{Z^{0}}^{2} \over m_{A^{0}}^{2}} 
   \tb \Delta_{\mu e}^{L} \\
   &\times  \left[ ( 4\cos^{2} \theta_{W}  { m_{Z^{0}}^{2} \over 
   m_{A^{0}}^{2}} - 2 ) F'(1) - 
   ( 2 \cos^{2} \theta_{W} - 1 ) {m_{Z^{0}}^{2} \over 
   m_{A^{0}}^{2}} \left( {1\over6} 
   \log{m_{A^{0}}^2 \over m_{Z^0}^2 } + {5\over18} \right) \right] \ .
\end{split} \label{bosonicapro3}
\end{equation}
Here, $F(\cos^2\theta_W) = 7.96$, $F'(\cos^2\theta_W) = 0.121$, and $F'(1) = 0.172$.

Notice that the diagram of $H^0$ including $W^-$ loop gives
contribution to $A^R$, which is proportional to $\tan\beta$ in a limit
of $m_{A^0}\gg m_{Z^0}$ and $\tan \beta \gg 1$. This is because the
$H^0W^+W^-$ coupling is suppressed by $\cos(\alpha-\beta)\sim
-2m_{Z^0}^2/(m_{A^0}^2\tan\beta)$ while the LFV Yukawa coupling in
Eq.~(\ref{Yukawa2}) is proportional to $\tan^2\beta$. All contributions 
from Barr--Zee diagrams to $A_R$ are proportional to $\tan\beta$ in a
limit of $m_{A^0}\gg m_{Z^0}$ and $\tan \beta \gg 1$. The exception is
those including the bottom-quark and tau-lepton loops, which tend to
be subdominant in moderate $\tan\beta$. It is argued in
Refs.~\cite{Paradisi:2006jp, Paradisi:2005tk} that the diagram of
$H^0$ including $W^-$ loop gives a contribution proportional to
$\tan^2\beta$ and that it is the largest among the Barr--Zee diagrams.
However, using corrected $\tb$ dependence, this $W^-$ loop
contribution is no longer the largest one, as will be shown below.

We also include Barr--Zee diagrams including $H^-$ loop, though the
contribution is also subdominant. It is also proportional to
$\tan\beta$ at most, and coefficient for its logarithmic term is
suppressed by $m_{Z^0}^2/m_{A^0}^2$.

\begin{figure}[h]
\begin{center}
   \begin{tabular}{cc}
      \includegraphics[width=7cm,clip]{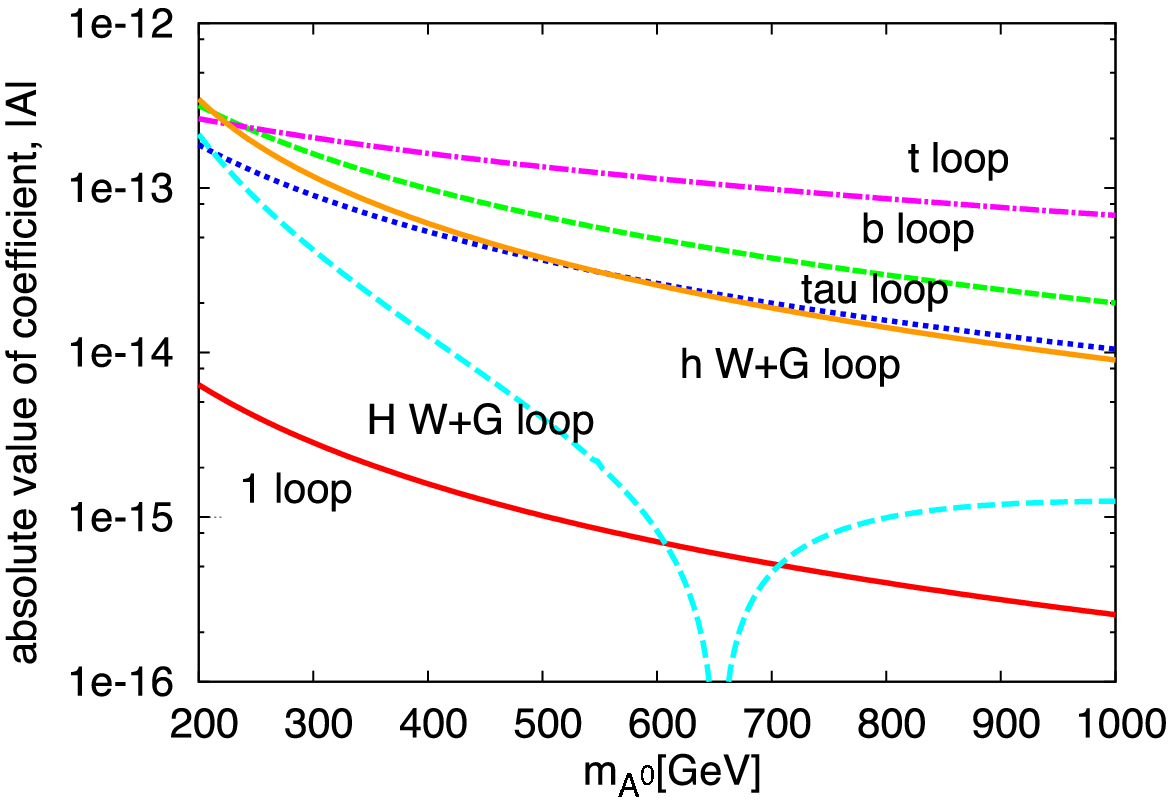} & 
       \includegraphics[width=7cm,clip]{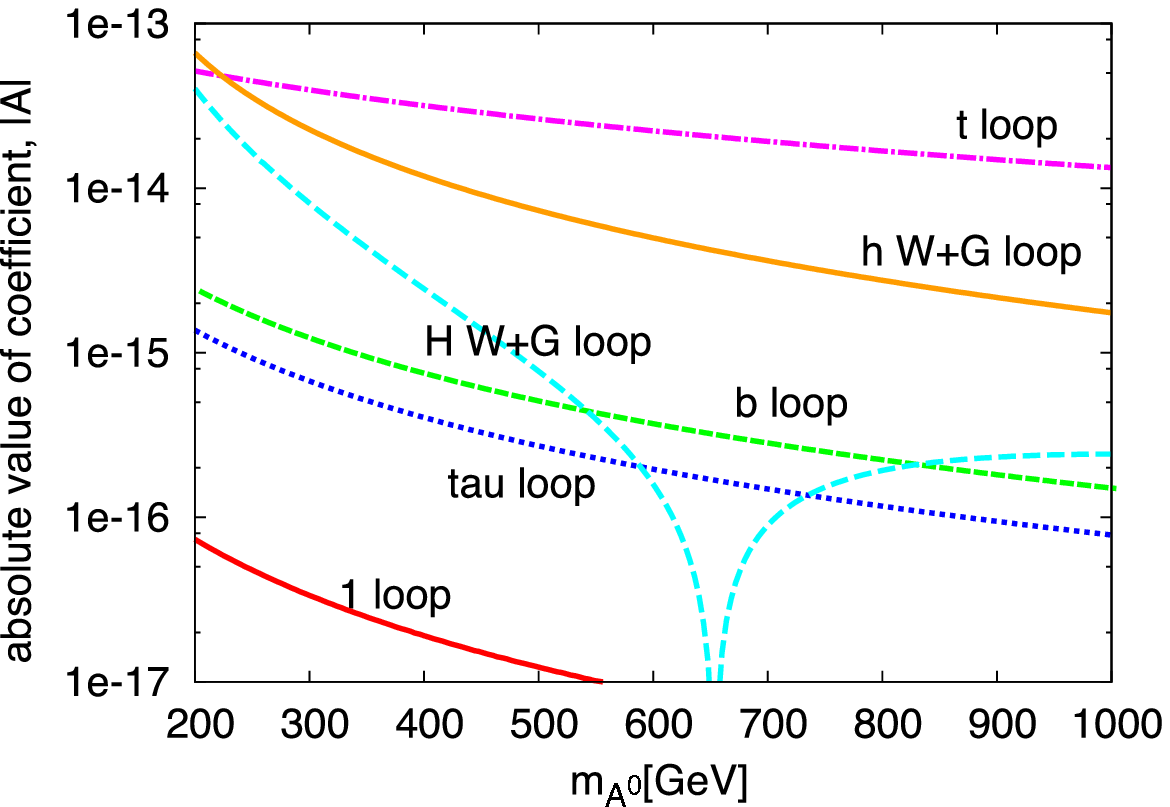} \\
   \end{tabular}
   \caption{Absolute values of coefficients for Higgs-induced dipole
     operator, $|A^R|$, as a function of CP-odd Higgs boson mass
     $m_{A^{0}}$.  We show those from diagrams including heavy-fermion
     loops, diagrams of light- and heavy-Higgs bosons including $W^-$
     and $G^-$ loops. For comparison, one-loop contribution to $A^R$
     is also shown. Here, left figure is for $\tb = 50$ and
     right one is for $\tb = 10$. We took $\Delta_{\mu e}^{L} = 5 \times
     10^{-6}$, $m_{h^0}=120$~GeV, $m_t(m_{Z^0})=181$~GeV, and $m_b(m_{Z^0})=3.0$~GeV.} 
\label{amplitude}
\end{center}
\end{figure}

Fig.~\ref{amplitude} shows each contribution to $A^R$ as a function of
CP-odd Higgs boson mass $m_{A^{0}}$ for $\tb = 50$ (left) and $\tb =
10$ (right).  Here, $\Delta_{\mu e}^{L} = 5 \times 10^{-6}$,
$m_{h^0}=120$~GeV, $m_t(m_{Z^0})=181$~GeV, and
$m_b(m_{Z^0})=3.0$~GeV. Barr--Zee diagrams including top-quark loop give
dominant contribution to $\mu\rightarrow e\gamma$, and the
bottom-quark one is also sizable only when $\tan\beta$ is large. The $W^-$
and NG-boson diagrams tend to be subdominant unless $m_{A^0}$ is
small. It is noticed in Refs.~\cite{Paradisi:2006jp, Paradisi:2005tk}
that there are strong cancellation between Barr--Zee diagrams of $H^0$
involving $W^-$ and $G^-$ loops \cite{Chang:1993kw}. However, other
contributions dominate over them, and hence this cancellation effect
does not appear in the branching ratio.

For comparison, we show the contribution from the one-loop diagrams 
(Fig.~\ref{1loop}) in Fig.~\ref{amplitude}. It is approximated as 
\begin{equation}
\begin{split}
   A^R_\text{one-loop} 
   &= { G_{F} \over 8\sqrt{2} \pi^{2} } ~  \Delta^{L}_{\mu e}
   \left[ 
   - ~ {\sin \alpha \cos(\alpha - \beta) \over \cos^{3}{\beta}} 
   {m_{\mu}^{2} \over m_{h^0}^2} ( {4 \over 3} 
   - \log { {m_{h^0}^2} \over m_\mu^2 } )  \right.    \\
   &+ ~ {\cos \alpha \sin(\alpha - \beta) \over \cos^{3}{\beta}} 
   {m_{\mu}^{2} \over m_{H^0}^2} ( {4 \over 3} 
   - \log{ {m_{H^0}^2} \over m_\mu^2 } ) 
   \left. + ~ {\sin \beta \over \cos^{3}{{\beta}}} 
   {m_{\mu}^{2} \over m_{A^0}^2} ( {5 \over 3} 
   - \log{ {m_{A^0}^2} \over m_\mu^2 } ) ~ 
   \right] \ .   \\
\end{split}
\end{equation}
As expected, this is always subdominant.

Now we consider competition between the gaugino- and Higgs-mediated
contributions to $\mu\rightarrow e \gamma$. The gaugino-mediated contribution 
to $A^R$ is approximated as 
\begin{equation}
\begin{split}
 A^{R}_{\rm gaugino} 
 = {1\over15} {\alpha_{2} \over 4 \pi} (1+{5\over4} \tan^{2} \theta_{W}) 
 {1\over \MSUSY^{2}} \delta^{LL}_{ji} \tb ,
\end{split}
\label{AL}
\end{equation}
where we take a common value $\MSUSY$ for the SUSY particle
masses. 

\begin{figure}[h]
\begin{center}
\includegraphics[width=9cm,clip ]{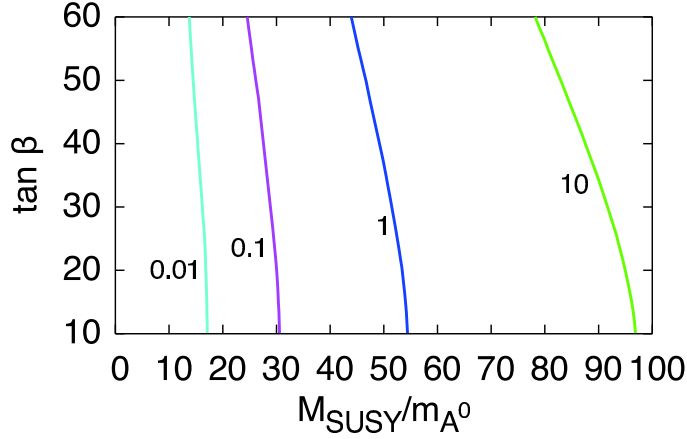} 
\caption{ Contour plot of square of ratio for Higgs- and
  gaugino-mediated contributions to $A^R$ as functions of $\MSUSY
  /m_{A^{0}}$ and $\tb$. Higgs-mediated contribution is dominant at
  right-handed side of linef on which this ratio is equal to unity.}
\label{competence}
\end{center}
\end{figure}

Fig.~\ref{competence} is a contour plot for square of ratio of the
Higgs- and gaugino-mediated contributions to $A^R$ as functions of
$\MSUSY / m_{A^{0}}$ and $\tb$.  The line on which this ratio is equal
to unity is boundary of the two regions where each effect dominates
$\meg$.  In small $\tb$ region, the Higgs-mediated contribution comes
from mainly Barr--Zee diagrams including top-quark loop. Both Higgs-
and gaugino-mediated contribution to $A_R$ are approximately
proportional to $\tan \beta$.  However, in large $\tb$ region, the
Barr--Zee diagrams with bottom-quark and tau-lepton loops receive
larger $\tan\beta$ enhancement, and the Higgs-mediated contribution
becomes larger with the same $\MSUSY/m_{A^{0}}$ value. We found that
the Higgs- and gaugino-mediated effects are comparable to each other
in $\meg$ when $\MSUSY / m_{A^{0}} \simeq 50 $.

%
\section{Correlation among LFV processes}

Now we discuss other $\mu$-$e$ transition processes, $\mu\rightarrow 3
e$ and $\mu$-$e$ conversion in nuclei, when the Higgs-mediated
contributions are dominant in the MSSM. These two processes have
strong correlation with $\meg$ when the gaugino-mediated contributions
are dominant, since effective LFV dipole operator determines the
processes.

First, we consider $\mu \rightarrow 3e$.  This process is generated
from three types of effective four-Fermi operators; scalar-, vector-,
and dipole operators. Scalar operators are induced by
tree-level Higgs boson exchange. On the other hand, vector and dipole
ones are generated by virtual-photon mediating processes $\meg^{*}$ at
higher order. When only the Higgs bosons contribute to $\mu \rightarrow 3e$,
vector operator mainly comes from one-loop diagrams, and
dipole one is generated by two-loop Barr--Zee diagrams. Since diagrams for
vector and scalar operators need two chirality flips, these operators
are suppressed by small Yukawa couplings ($y_{\mu}$  or $y_{e}$),
compared to dipole operator. Vector and dipole operators come from
higher-order effects and they are suppressed by loop factors.

Thus, contributions of these operators are estimated roughly as follows,
\begin{equation}
\begin{split}
   & A_{0} \simeq {y_{\mu} y_{e} \Delta^{L}_{\mu e} 
   \over m_{A^0}^{2} } \tan^{3} \beta\ , \\
   & A_{1} \simeq {\alpha_{\rm em} \over 4 \pi}  
   {y_{\mu}^{2} \Delta^{L}_{\mu e} \over m_{A^0}^{2} } 
   \tan^{3} \beta \log \bigg({m_{\mu}^{2} 
   \over m_{A^0}^{2}} \bigg)\ ,  \\ 
   & A_{2} \simeq \left( {\alpha_{\rm em} \over 4 \pi } \right)^{2} 
   {y_{t}^{2} \Delta^{L}_{\mu e} \over m_{A^0}^{2}} \tan\beta 
   \left[ \log \bigg({m_{t}^{2} \over m_{A^0}^{2} } \bigg) \right]^{2}\ , 
\end{split}
\end{equation}
where lower indices (0,1,2) mean coefficients for scalar, vector, and
dipole operators, respectively. Ratio of these coefficients becomes
$A_{0} : A_{1} : A_{2} \simeq 1 : O(1) : O(10)$, and the
coefficient for dipole operator
is the dominant contribution. There is also
$\log \left({ m_{\mu}^{2} /m_{e}^{2} } \right) $ enhancement for
dipole operator contribution to $\mu \rightarrow 3e$, which comes from
final state phase space integral.  As a consequence, $\meee$
is dominated by dipole operator and there is strong correlation
between $\meee$ and $\meg$,
\begin{equation}
\frac{\BR(\mu \rightarrow 3e)}{\BR(\meg)} 
\simeq {\alpha_{\rm em} \over 3\pi} \left( \log \left({ m_{\mu}^{2} \over m_{e}^{2} } \right) 
- {11\over4} \right) \simeq 0.006.
\end{equation}


Next, we discuss $\mu$-$e$ conversion in nuclei. Dominant contribution
to this process comes from tree-level Higgs boson exchange
\cite{Kitano:2003wn}. The reason is that coupling between Higgs
boson and nucleon is characterized by the nucleon mass $m_N$ through
the conformal anomaly relation~\cite{Shifman:1978zn}, and could evade 
suppression of light constituent quark mass. Then,
branching ratio for $\mu$-$e$ conversion in nuclei at large $\tb$ is derived 
from formulae in Refs.~\cite{Kitano:2002mt,Cirigliano:2009bz} as follows, 
\begin{equation}
\BR (\mu {\rm Al} \rightarrow e {\rm Al}) \simeq
6.8\times 10^{-5}\,
\frac{G_F^2 m^{7}_{\mu}m^{2}_{p}}{m^{4}_{H^{0}}\omega^{\rm Al}_{\rm capt}}\,
( \Delta^{L}_{\mu e})^{2} \tan^{6} {\beta} \,.
\end{equation}
Here, $\omega^{\rm Al}_{\rm capt}\simeq 0.7054 \times 10^{6}
\text{sec}^{-1}$, and $m_{p}$ is proton mass.  We use the recent
lattice simulation result \cite{Ohki:2008ff} for the $\sigma$ term,
which shows that the strange quark content of the nucleon is much
smaller than previously thought. Notice that branching ratio for
$\mu$-$e$ conversion in nuclei is scaled by $\tan^6\beta$, while those
for $\meg$ and $\meee$ are proportional to $\tan^2\beta$.

\begin{figure}[h]
\begin{center}
   \begin{tabular}{cc}
\includegraphics[width=7cm,clip ]{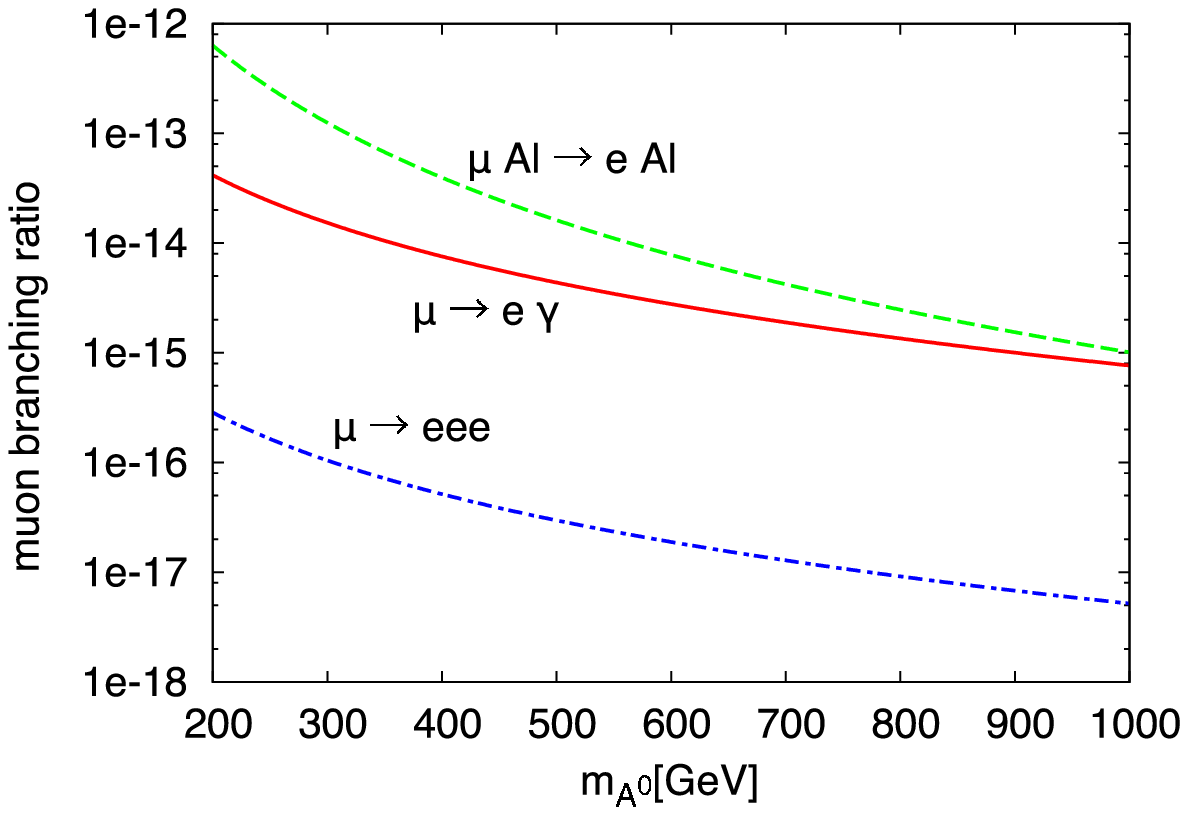}  &
\includegraphics[width=7cm,clip ]{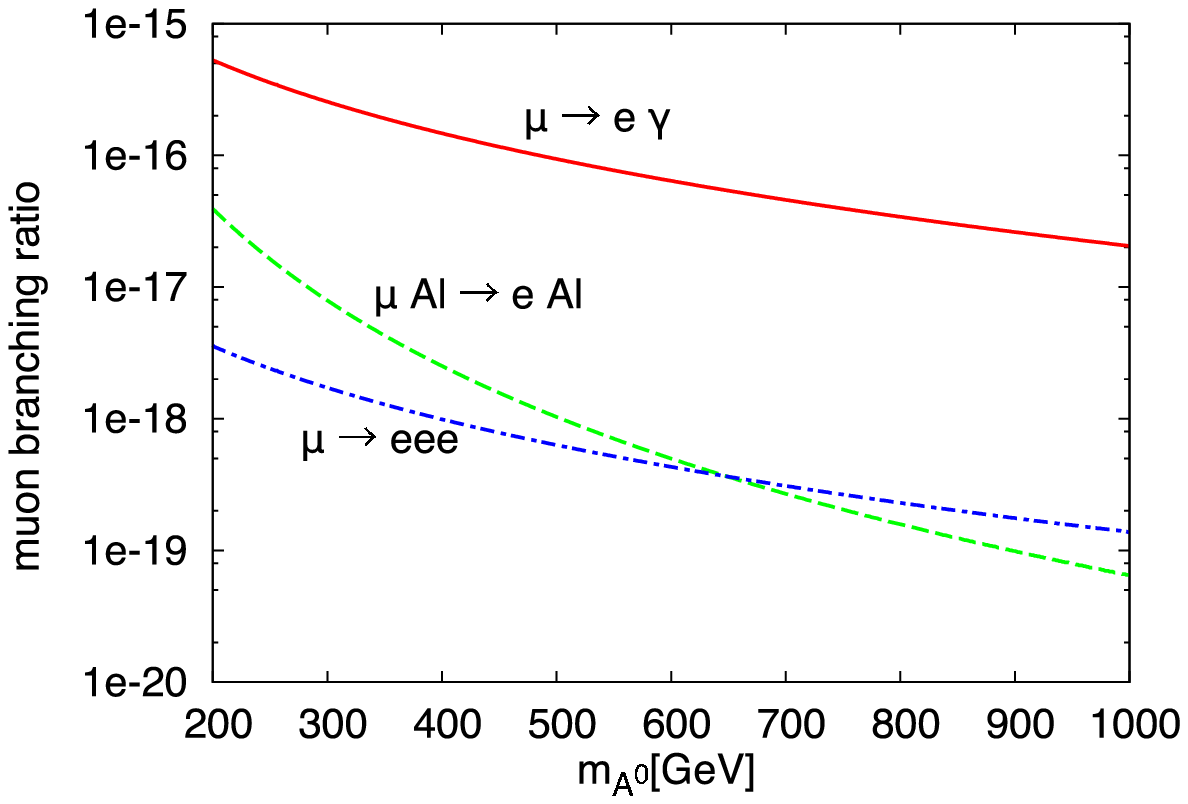}  \\
\end{tabular}
\caption{Branching ratios of Higgs-mediated cLFV processes. We took
  $\tb = 50$ (left) and $\tb = 10$ (right), $\Delta^{L}_{\mu e} = 5.0
  \times 10^{-6}$. }
\label{HiggsLFV}
\end{center}
\end{figure}%

In Fig.~\ref{HiggsLFV} branching ratios of Higgs-mediated LFV
processes are shown as a function of $m_{A^0}$. Though we include
contributions from the scalar and vector operators in the evaluation
of $\BR(\meee)$ in addition to the dipole one, it is found from this
figure that there is still tight correlation between $\meg$ and $\mu
\rightarrow 3e$, ${\BR(\meg)/\BR(\meee) } \simeq O(\alpha_{\rm em}) $.
Thus, it is a signature that dipole operator dominates these two
processes.  On the other hand, $\mu$-$e$ conversion in nuclei is
dominated by tree-level Higgs boson exchange, and such simple
correlation does not appear, as expected. In the gaugino-mediation
case, the dipole operator dominates three processes. Thus, it is
important to measure $\mu$-$e$ conversion rate for discrimination of
these two cases, in addition to $\meg$.

It is also found that while $\mu$-$e$ conversion process is simply
scaled as $1/m_{A^0}^4$, other two processes are not. This is because
other two processes receive large logarithmic corrections.

\begin{figure}[h]
\begin{center}
   \includegraphics[width=9cm,clip ]{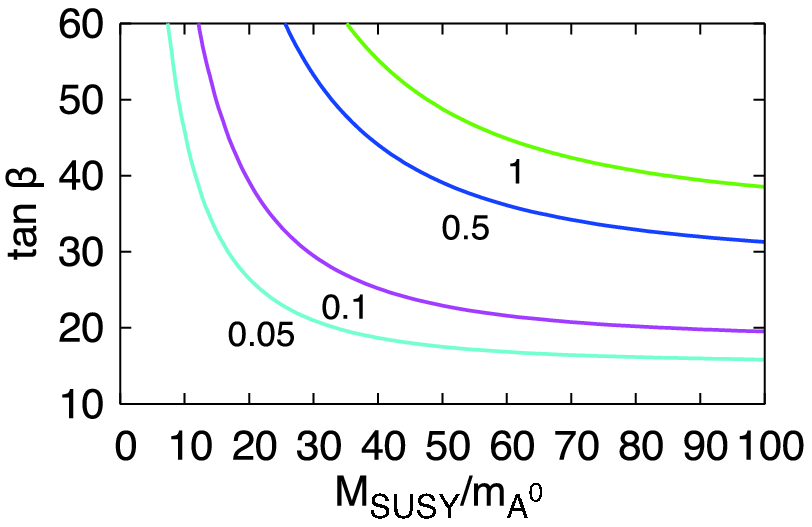} 
   \caption{Contour plot of BR($\mu {\rm Al} \rightarrow e {\rm Al}$)
     / BR($\meg$), $\tb$ vs $\MSUSY / m_{A^{0}}$ including both the
     Higgs- and gaugino-mediated contributions. }
   \label{observation}
\end{center}
\end{figure}%

Fig.~\ref{observation} shows contour plot of BR($\mu {\rm Al}
\rightarrow e {\rm Al}$) / BR($\meg$) including both the Higgs- and
gaugino-mediated contributions. If the Higgs-mediated contribution is
dominant in the cLFV processes, the ratio between $\meg$ and $\mNeN$ is
sensitive to $\tan\beta$, but not to $\MSUSY/m_{A^0}$. On the other
hand, if gaugino-mediated LFV is dominant, this ratio is about $O(
\alpha_{\rm em} ) $ since dipole operator contributions dominate the
cLFV processes. When $\MSUSY/m_{A^0}\sim (10-50)$ and $\tan\beta\gsim
10$, both Higgs- and gaugino-mediated diagrams contribute to those
processes in different way and we could give constraints $\MSUSY /
m_{A^{0}}$ and $\tan\beta$ from $\BR(\mu {\rm Al} \rightarrow e {\rm
  Al}) / \BR(\meg)$ .

\section{Conclusions and discussion} 

In this letter, we reevaluated $\mu$-$e$ transition processes induced
by non-holomorphic Yukawa interactions in the MSSM.
We discussed
correlation among branching ratios for $\mu \to e \gamma$, $\mu \to 3
e$, and $\mu$-$e$ conversion in nuclei in the MSSM, by including
both the gaugino- and Higgs-mediated contributions to the
processes. It was assumed in this letter that left-handed sleptons
have flavor-mixing mass terms.

Though Higgs-mediated contribution to $\mu$-$e$ transition processes is
evaluated in previous works~\cite{Paradisi:2006jp, Paradisi:2005tk},
we found that contribution from Barr--Zee diagram including $W^-$
boson, which was thought to be the largest contribution to $\meg$
among various Higgs-mediated contributions, has incorrect dependence
on $\tan\beta$. As a result, branching ratio for $\meg$ was
overestimated. We showed that Barr--Zee diagrams including top quark
are rather dominant, and those including bottom quark and tau lepton
are also sizable only when $\tan\beta$ is large. Then, the Higgs-mediated
contribution dominates over the gaugino-mediated one in
$\mu\rightarrow e \gamma$ when $\MSUSY/m_{A^0}\gsim 50 $,
irrespectively of $\tan\beta$ as far as $\tan\beta$ is not large.

We evaluated ratio of branching ratios for $\mu\rightarrow e \gamma$
and $\mu$-$e$ conversion in nuclei by including both the gaugino- and
Higgs-mediated contributions to the processes. We found that the ratio
is sensitive to $\tan\beta$ and $\MSUSY/m_{A^0}$ when
$\MSUSY/m_{A^0}\sim (10-50)$ and $\tan\beta\gsim 10$. Ratio of the
branching ratios for $\mu\rightarrow e \gamma$ and $\mu\rightarrow 3
e$ is insensitive to $\tan\beta$ and the MSSM mass spectrum, since the
dipole term contribution is always dominant in $\mu\rightarrow 3 e$.

In general, right-handed slepton mass terms or $A$ terms could be
sources for flavor-mixing. In particular, gaugino-mediated
contributions from right-handed slepton mass receive destructive
interference between the bino and bino-Higgsino amplitudes
\cite{Hisano:1996qq}.  Therefore, Higgs-mediated contribution may be
significant in some parameter region and decoupling behavior of
$\MSUSY$ could be modified.  We leave it for our future work.

\appendix

\section{Barr--Zee integrals}
The Barr--Zee integrals $f(z)$, $g(z)$ and $h(z)$ are given
by
\begin{equation}
\begin{split}
   f(z) 
   &= {1 \over 2} z \int_{0}^{1} dx 
   { 1-2x(1-x) \over x(1-x) -z } \log{x(1-x) \over z} \ ,\\
   g(z) 
   &=  {1 \over 2} z \int_{0}^{1} dx 
   { 1 \over x(1-x) -z } \log{x(1-x) \over z} \ ,  \\ 
   h(z)
   &\left(= z^2 {d \over dz}\biggl({g(z) \over z}\biggr)\right)  \\ 
   & ={z\over2}\int_0^1 {dx\over z-x(1-x)} 
         \biggl[1+{z\over z-x(1-x)}\log{x(1-x)\over z} \biggr] \ .
\end{split}   
\end{equation}
In the limit of $1 \gg z$, the asymptotic forms of them is given as follows~\cite{Chang:1993kw},  
\begin{equation}
\begin{split}
&   f(z) \sim {z\over 2} (\log z)^{2} \ ,~   
   g(z) \sim {z\over 2} (\log z)^{2} \ , ~     
   h(z) \sim ~ z \log z \ , ~ \\
&   f(z) - g(z)  \sim z ( \log z + 2) \ . 
 \end{split}
 \end{equation}
On the other hand, in the limit of $1 \ll z$, 
\begin{equation}
\begin{split}
 &  f(z) \sim {1 \over 3} \log z+ {13 \over 18} ,~  
   g(z) \sim {1 \over 2} \log z+ 1 ,~
   h(z) \sim -{1\over2} (\log z +1) \ , ~\\
 &  f(z) - g(z) \sim - {1 \over 6}  \log z - {5 \over 18} \ . 
 \end{split}
 \end{equation}

 Similarly, in the limit of $1 \gg z$, the asymptotic forms of $F(z) = 3 f(z) + 5 g(z) + {3/4} g(z)+{3/4} h(z)$ and
$F'(z) = (g(z) -f (z))/(2z)$ are derived as 
\begin{equation}
\begin{split}
F(z) 
&\sim {35\over8} z (\log z)^{2} + {3\over4} z \log z \ , \\
F'(z) &\sim -{1\over2} (\log z + 2) \ .
\end{split}
\end{equation}
On the other hand, in the limit of $1 \ll z$, 
\begin{equation}
\begin{split}
F(z) 
& \sim {7\over2} \log z + {181 \over 24} \ , \\
F'(z) &\sim {1 \over 2z} ({1 \over 6} \log z + {5 \over 18}) \  . 
\end{split}
\end{equation}

\section*{Acknowledgment}

The work was supported in part by the Grant-in-Aid for the Ministry of
Education, Culture, Sports, Science, and Technology, Government of
Japan, No. 20244037, No. 2054252, No. 2244021 (J.H.) and No. 
20007555 (M.Y.). The work of J.H. is also supported by the World 
Premier International Research Center Initiative (WPI Initiative), MEXT, 
Japan. This work was supported by Maskawa Institute in Kyoto Sangyo 
University (M.Y.). The work of S. S. was financially supported by the Sasakawa 
 Scientific Research Grant from The Japan Science Society.


{}

\end{document}